\documentclass[11pt]{article}
\usepackage{graphicx}
\usepackage{amssymb}

\begin{document}

\title{  Nonpartonic effects in pion electroproduction \\
in the HERMES kinematical region  }

   \author{V. Uleshchenko$^{1}$ and A. Szczurek$^{2,3}$  \\
   {\it $^{1}$ Institute for Nuclear Research, 03-680 Kyiv, Ukraine  } \\
   {\it $^{2}$ Institute of Nuclear Physics, PL-31-342 Cracow, Poland  } \\
   {\it $^{3}$ University of Rzesz\'ow, PL-35-959 Rzesz\'ow, Poland }  }

\maketitle

\begin{abstract}
The presentation is concerned with higher twist corrections of nonpartonic
origin to semi-inclusive observables in the kinematical region
relevant for the HERMES experiment.
We demonstrate a strong impact of the VDM-like interaction and
the exclusive production of $\rho^0$ meson
on the extraction of the $\bar d - \bar u$ asymmetry from charged pion
DIS multiplicities.
We also show that it is the exclusive-$\rho^0$ channel which accounts
for the experimentally observed effect of the excess of
charged over neutral pions produced at large $z$.
\end{abstract}

\section{Introduction}

Recently the interest in semi-inclusive processes has increased
significantly. They appear to be a very useful tool
for separating the flavour- and spin-dependent quark distributions
in the nucleon.
The interpretation of such experiments is
based on the quark-parton model, often in its simplest form.
Therefore, from a practical point of view, some of very important
topics are: the presence of mechanisms
beyond the QPM in the studied reactions, contributions of such mechanisms
to measured cross sections and their influence on the analysed quantities.

The HERMES Collaboration has precisely measured multiplicities of
charged and neutral pions produced in unpolarized DIS off the proton
\cite{HRM_pions}. Previously, charged pion production
off proton and deutron targets was used for
the extraction of the $\bar d - \bar u$ asymmetry \cite{HRM_db_ub}.
Since the HERMES expe-riment corresponds to the kinematical region of
relatively small momentum transfer
where the applicability of the quark-parton model is not obvious
a possible influence of mechanisms beyond QPM on the
HERMES observables should be of particular interest.

\section{Effects in the extraction of the $\bar d- \bar u$ asymmetry}

An extraction of the light quark asymmetry of the nucleon sea
from charged pion multiplicities performed by HERMES \cite{HRM_db_ub}
is based on the pure QPM description of the pion electroproduction.
The pion production process is assumed to take place in two steps: first,
a hard interaction of the incoming virtual photon with a quark in the
nucleon, and second, a hadronisation of the stuck quark into final pion(s).
Then, one can combine yields of positive and
negative pions produced off proton and neutron, and using isospin
invariance, isolate a quantity sensitive to the
$\bar d - \bar u$ asymmetry \cite{HRM_db_ub}:
\begin{equation}
\frac{ \bar d(x) - \bar u(x) }{ u(x) - d(x) } =
\frac{J(z) [1 - r(x,z)] - [1 + r(x,z)]}
     {J(z) [1 - r(x,z)] + [1 + r(x,z)]} \; ,
\label{db_ub_extract}
\end{equation}
where \
$
J(z) = \frac{3}{5} \cdot
\frac{1 + D_{-}(z)/D_{+}(z)}{1 - D_{-}(z)/D_{+}(z)}
$ \
depends on the favoured $D_+(z)$ and unfavoured $D_-(z)$
[light quark]-pion fragmentation functions,
and \
$
r(x,z) = $ $ \frac{N_p^{\pi^-}(x,z) - N_n^{\pi^-}(x,z)}
                   {N_p^{\pi^+}(x,z) - N_n^{\pi^+}(x,z)}
$ \
is a ratio of differences of charged pion yields off proton and neutron.

In such an approach pions produced via other than partonic mechanisms
are ignored.
At the same time,
{\em  experimental\/} multiplicities $N_{p,n}^{\pi^+,\pi^-}$, which
one inputs into the right-hand side of Eq.(\ref{db_ub_extract})
in order to obtain expe-rimental $\bar d - \bar u$ asymmetry,
{\em do\/} contain such nonpartonically produced pions. Thus,
Eq.(\ref{db_ub_extract}) is not an exact equation. In the $Q^2 \to \infty$
limit the nonpartonic contributions vanish, but in the HERMES kinematical
region where
$\langle Q^2 \rangle \approx 2.5$~GeV$^2$ they cannot be neglected.
Two mechanisms appear to be especially important: so-called
VDM-like production and exclusive-$\rho^0$ channel \cite{SUS_db_ub}.
In the VDM-like production the incoming virtual photon interacts with the
nucleon strongly, via its intermediate vector meson state
into which it fluctuates just before the interaction.
The exclusive-$\rho^0$ channel corresponds to the reaction
$\gamma^* + N \to \rho^0 + N$ with a subsequent decay
$\rho^0 \to \pi^- \pi^+$.

Based on the world data for pion production in $\pi N$ collisions and for
$\rho^0$ electroproduction we have estimated contributions of these
mechanisms%
\footnote{ For a detailed analysis see Ref.\cite{SUS_db_ub}. }
and found them noteworthy.

\begin{figure}
  \begin{center}
    \includegraphics[width=8.7cm]{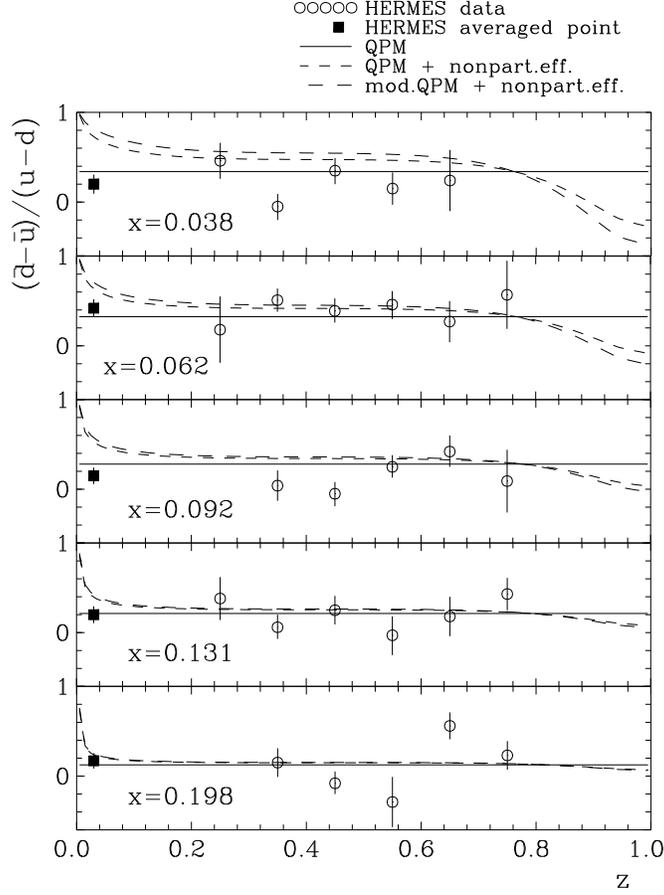}
  \end{center}
\caption{\small
$\frac{\bar d - \bar u}{u-d}$ obtained from the r.h.s.\ of
Eq.(\ref{db_ub_extract}) in different bins of Bjorken-$x$.
}
\label{Hermes_z}
\end{figure}

The influence of the VMD and
exclusive-$\rho^0$ channels on the measured quantity
$\frac{\bar d - \bar u}{u-d}$ can be seen in Fig.\ref{Hermes_z}.
The shown HERMES points were obtained
with experimentally measured multiplicities $N_{p,n}^{\pi^+,\pi^-}$.
The solid curve independent of $z$ represents the
pure QPM result calculable from both sides of Eq.(\ref{db_ub_extract}).
The two dashed curves correspond
to calculations in which contributions of the nonpartonic mechanisms
are included in two ways: either by simple adding them to the main
partonic term, or with additional modification of the QPM term by a damping
$Q^2$-dependent factor which is required for QPM to work properly at such a
low $Q^2$ \cite{SU1_2}.
A discrepancy between the solid and dashed lines gives us a theoretical
error on the working formula~(\ref{db_ub_extract}). As one can see, it is
comparable to or even larger then the given {\em total\/}
(containing both statistical and systematic components) uncertainty of the
final $z$-averaged HERMES experimental points (solid squares).
The effect is stronger for the case of modified QPM term, i.e.\ for the
more correct one. The averaging in
$z$ also looks doubtful because of the clearly visible $z$-dependence.

\section{Excess of charged over neutral pions at large $z$}

A very interesting phenomenon has been recently observed
by the HERMES Collaboration:
different yields of charged and neutral pions produced off proton
at large $z$:
$\frac{1}{2} ( N_p^{\pi^+} \! + N_p^{\pi^-} ) > N_p^{\pi^0}$
for $z \gtrsim 0.7$
\cite{HRM_pions}.
\begin{figure}
  \begin{center}
    \includegraphics[width=6.9cm]{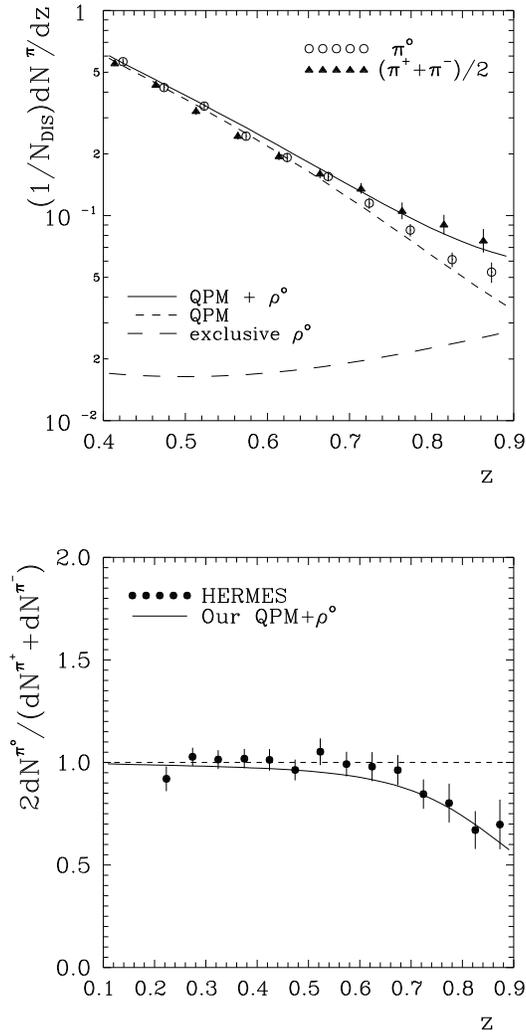}
  \end{center}
\caption{\small
Neutral and charged pion multiplicities (top panel) and
their ratio (bottom panel) as a function of $z$.
}
\label{ht_our}
\end{figure}
In the QPM picture with quark fragmentation there is no room for such an
effect; there the neutral pion yield is exactly equal, due to isospin
symmetry, to the charged pion yield. As a possible source for the
difference, alongside the instrumental resolution, exclusive pion
production channels (e.g.\ $\gamma^* + p \to \pi^+ +\Delta^0$) were
suggested \cite{HRM_pions}.  In our opinion however, exclusive
{\em pion\/}-production
channels cannot be responsible for an effect in a broad region
of $z$ since pions produced in this way
carry almost the whole energy of the virtual photon and therefore
contribute to measured multiplicities only
in a narrow region near $z \sim 1$.
In contrast, the discussed in the previous section
the exclusive production of the
$\rho^0$ meson, which decays into two pions, gives a contribution to the
charged pions multiplicity in a broad $z$-range.
It is especially important at larger $z$ where the partonic rate is smaller.
The contributions
to the neutral pion multiplicity, caused e.g.\ by the exclusive-$\rho^\pm$
channels, are much smaller due to smaller cross sections of these processes
($\rho^0$ production, known as the dominant exclusive $\gamma^* N$ channel,
is dominated by the Pomeron exchange which is absent in the case of
charged-$\rho$ channels).

In the top panel of Fig.\ref{ht_our} we show the pion multiplicities
for $z > 0.4$. The experimental points are from \cite{HRM_pions}.
The short-dashed curve represents the QPM
prediction which is the same for both types of multiplicities. The solid
curve includes also the contribution of the exclusive-$\rho^0$
channel to the charged pion yield
(this contribution is also shown separately).
As one can see, the exclusive-$\rho^0$
channel nicely explains the observed effect. This is especially well visible
in the bottom panel where the ratio of charged-to-neutral pion
multiplicities is shown.

\section{Conclusions}

We have discussed the role of some mechanisms beyond the quark-parton
model in semi-inclusive production of pions in the kinematicl region
relevant for the HERMES experiment.

We have shown that production of pions via VDM-like interaction and via
exclusive-$\rho^0$ channel have a strong impact on the extraction of the
$\bar d - \bar u$ asymmetry from the charged
pion DIS multiplicities which was done by the HERMES Collaboration.
The extracted asymmetry turns out to be very sensitive to
the influence of the mentioned mechanisms and
neglecting them leads to a significantly distorted result.

We have also shown that the exclusive production of $\rho^0$ mesons
accounts for the experimentally observed excess of charged over
neutral pions produced at large $z$.

To conclude, nonpartonic mechanisms
are essential for a good theore-tical description of the pion
electroproduction in the HERMES kinematical region and
cannot be omitted in a proper analysis.
In addidition, we would like to stress that
similar nonpartonic correction should be expected in polarised processes,
as well.



\begin{thebibliography}{99}

\bibitem{HRM_pions}
HERMES Collaboration, A. Airapetian et al.,
Eur. Phys. J. {\bf C21} (2001) 599.

\bibitem{HRM_db_ub}
HERMES Collaboration, K. Ackerstaff et al.,
Phys. Rev. Lett. {\bf 81} (1998) 5519.

\bibitem{SUS_db_ub}
A. Szczurek, V. Uleshchenko and J. Speth,
Phys. Rev. {\bf D63} (2001) 114005.

\bibitem{SU1_2}
A. Szczurek and V. Uleshchenko,
Eur. Phys. J. {\bf C12} (2000) 663; \\
A. Szczurek and V. Uleshchenko,
Phys. Lett. {\bf B475} (2000) 120.

\end{thebibliography}
\end{document}